\def\astrobj#1{#1}
\documentclass[]{elsart}
\usepackage{graphicx}
\usepackage{amssymb}
\begin{document}
\begin{frontmatter}

   \title{The OH 1612\,MHz maser pump rates of stellar, interstellar and post-AGB OH masers.}

   \author{J. H. He} 
   \ead{mailhejh@yahoo.com.cn}
   \address{National Astronomical Observatories/Yunnan Observatory, Chinese Academy of Sciences, 
                  P.O. Box 110, Kunming, 650011, PR China}

   \begin{abstract}

(Pseudo) radiative pumprate of OH 1612\,MHz masers is defined for a sample of 44 OH/IR sources 
(infrared sources with OH 1612\,MHz maser), irrespective of the real maser pumping mechanisms. 
The correlation between the (pseudo) maser pumprates and the evolutionary status of the maser 
sources reveals that the radiative pumprates of stellar OH masers are nearly fixed, which agrees with the 
theoretical prediction for radiatively pumped OH maser. The (pseudo) radiative pumprates of interstellar 
OH masers are not only very small but also varying broadly over two orders of magnitude, which is argued to 
be the manifestation of varying number of quiet absorbing OH cloudlets 
and/or various OH maser pumping mechanisms 
and/or competitive gain between mainline and 1612\,MHz OH masers 
and/or anisotropy of the maser emission. The 
radiative pumprates of post-AGB OH masers very possibly decrease with increasing IRAS C$_{32}$ 
color indices and distribute in an interim region between the stellar and interstellar OH masers in the 
pumprate-color diagram.

\end{abstract}


\end{frontmatter}

\section{Introduction}

OH 1612\,MHz masers have been found in different kinds of objects, such as OH Miras, OH/IR stars, 
Proto-Planetary Nebulae (PPNe), Planetary Nebulae (PNe), Red Supergiants (RSGs), H\,II regions
(H\,IIs), molecular clouds, and so on (Elitzur,\,\cite{eli92}). The OH masers appearing in H\,IIs or 
molecular clouds, are called {\it interstellar OH masers} while those appearing in all other types of sources are 
called {\it stellar OH masers}, depending on whether the masing molecules originate from interstellar space (by accretion) 
or from the central star (by stellar wind). Because the OH masers in post-AGB sources, such as post-AGB stars, 
PPNe and PNe, show peculiar properties, as we will see below in the discussions of this paper, I would like to separate them 
from the stellar OH maser group to form a third group: {\it post-AGB OH masers}. Up till now, three different pumping 
mechanisms have been proposed for the OH 1612\,MHz masers arising in different astronomical environments: 
radiative pumping, collisional pumping and chemical 
pumping. According to the recent developments in the maser pumping theory, more physical effects such 
as local and non-local line overlap (e.g., Pavlakis and Kylafis\,\cite{pav96a}, \cite{pav96b}) were recognized 
to be playing important roles in certain cases. But little effort has been made to directly compare in observations 
the different pumping mechanisms of the OH 1612\,MHz masers that occur in different kind of objects.

Among the different kinds of objects showing OH 1612\,MHz maser, late type stars such as OH Miras, OH/IR stars 
and RSGs are believed to have their OH masers radiatively pumped. Most of these objects have a huge 
spherical, although perhaps clumpy, circumstellar dust-gas envelope. The OH 1612\,MHz maser is formed in 
the outer part of the huge envelope and is pumped mainly by absorbing the 34.6 and 53.3\,$\mu$m photons. 
The physical environments in the three kinds of objects are a little different. For example, OH Miras often have 
optically thin dust envelope while OH/IR stars are usually embedded in very optically thick dust shell; 
RSGs have much higher luminosity and their dusty envelopes are harsher than that of OH Miras and 
OH/IR stars. 

PPNe and PNe are objects immediately following the AGB evolutionary stage. The OH 1612\,MHz maser 
inherited from their precursors can still survive for about 1000 years (Sun and Kwok\,\cite{sun87}). But 
due to the evident changes of the envelope, the survived OH 1612\,MHz masers in PPNe or PNe are often 
found not in a spherical shell but in a circumstellar disk or bipolar outflow (e.g., Zijlstra et al.\,\cite{zij01}). 
The pump mechanism of these remanent OH masers is thought to be mainly radiative one too. However, 
for PPNe or PNe that show bipolar outflows, the OH masers can also appear in the interaction region 
between the outflow and the ambient interstellar material (Likkel and Morris\,\cite{lik88}) and, in this case, 
the OH masers may be pumped by collision.

OH 1612\,MHz masers have been found in some H\,IIs in Star formation Regions (SFRs). 
They are often found accompanying with the OH main line masers (Caswell\,\cite{cas99}, Szymczak and 
Gerard\,\cite{szy04a}). The physical environment in SFRs is much more complicated than that in the circumstellar 
envelope of late type stars. The expanding ionized hydrogen sphere creates a geometrically thin but dense 
layer compressed by shock waves in the inner edge of the huge outer in-falling dust-gas envelope. The OH masers were 
believed to arise just in the shock compressed layer (Cochran and Ostriker\,\cite{coc77}). Additionally, high 
velocity bipolar outflow have been found in many SFRs (e.g., Campbell\,\cite{cam86}, Shepherd\,\cite{she96}), 
but it is not clear whether the OH 1612\,MHz masers can be stimulated in the interaction region between the 
outflow and the ambient interstellar material. 

The difference in physical environments results in different OH maser-pumping mechanisms at work. 
However, it is not easy to directly discriminate the different maser pumping mechanisms from observational 
aspects. In our previous two papers (He et al.\,\cite{he04a} and He \& Chen \cite{he04b}, hereafter 
Paper I and Paper II, respectively), we systematically analyzed the 34.6 and 53.3\,$\mu$m pumping lines in all 
Infrared Space Observatory (ISO) Short Wavelength Spectrometer (SWS) and Long Wavelength Spectrometer 
(LWS) spectra of known OH/IR sources (infrared sources showing OH 1612\,MHz maser) and found that the 
(pseudo) radiative pumprates of OH masers arising in different environments are quite different. This paper 
continues to discuss another important problem: how the different kinds of 1612\,MHz maser sources obey or 
violate the principle that the input of pumping photons should be proportional to the maser output for a radiatively 
pumped maser? Although similar job had been done in the past decades by other authors using IR photometrical 
data (e.g., Le Bertre et al.\,\cite{leb84}, Dickinson\,\cite{dic87}, Silva et al.\,\cite{sil93}), 
the available IR spectral data from ISO enable one to perform a more powerful analysis of the OH 1612\,MHz maser 
pump rate by directly comparing the emission of maser photons to the absorption of IR photons. The potential 
of ISO spectra is dug out to the utmost to constrain the value or a reasonable range of the (pseudo) radiative 
maser pumprate for all OH/IR sample sources. In section\,\ref{data}, the data sets are organized and 
described. Some sources with ambiguous source types are re-identified based on literature information in 
section\,\ref{identification}. The discussion of the (pseudo) radiative pumprates of different types of OH maser is given in 
section\,\ref{pumprate}. Some related topics are discussed in Section\,\ref{discussion}. Section\,\ref{summary} 
briefly summarizes the main conclusions.


\section{Observational data}
\label{data}

Although not all OH 1612\,MHz masers considered here are radiatively pumped, a {\it pseudo radiative pumprate} 
can be defined by temporarily assuming they are all radiatively pumped. This pseudo pumprate is defined as the 
integrated 1612\,MHz OH maser photon flux divided by the sum of the integrated infrared absorption line photon 
fluxes at both wavelength: 34.6 and 53.3\,$\mu$m. As we demonstrated in paper I and II, this (pseudo) pumprate 
has the advantage to effectively distinguish OH masers arising in different physical environments and/or pumped 
by different mechanisms. 

According to Paper I and II, six OH/IR sources have both the 34.6 and 53.3\,$\mu$m pumping lines detected by 
ISO; their (pseudo) maser pumprates that is defined as integrated line photon flux ratio 
$n_{\rm OH}/(n_{34.5}+n_{53.3})$ had already been derived therein. Here the pumprates 
of the six sources are reproduced in Table\,\ref{sample1} and these sources compose the `Sample 1' of this paper. Also 
given in the table are the source type mainly given by SIMBAD database (http://simbad.u-strasbg.fr/sim-fid.pl) and 
the IRAS color C$_{32}$=log({\it F}$_{60}$/{\it F}$_{25}$), where in the case that no reliable IRAS fluxes 
are available for defining the color indices ISO spectra are used to derive equivalent IRAS fluxes using the 
IRAS photometrical filter templates in ISAP.\footnote{The ISO Spectral Analysis Package (ISAP) is a joint 
development by the LWS and SWS Instrument Teams and 
Data Centers. Contributing institutes are CESR, IAS, IPAC, MPE, RAL and SRON.} (More remarks are given 
upon the derivation of equivalent IRAS fluxes in the last paragraph of this section.)
%
\begin{table}[]{}
\scriptsize
\caption[]{\scriptsize Sample one: objects with both OH-maser infrared-pumping lines detected.}
\label{sample1}
\centering
\begin{tabular}{l@{ }l@{  }l@{  }l@{  }}
\hline
\noalign{\smallskip}
Name            &Type    &C$_{32}$  &Pumprate$^\dagger$            \\
(1)             &(2)     &(3)       &(4)                \\
\noalign{\smallskip}
\hline
\noalign{\smallskip}
\astrobj{07209$-$2540}      &RSG               &-0.66     &0.079              \\
\astrobj{17424$-$2859}      &\astrobj{Sgr A*}  &0.56      &3.56e-6            \\
\astrobj{17441$-$2822}      &\astrobj{Sgr B2}  &2.28      &1.56e-5            \\
\astrobj{19244$+$1115}      &RSG               &-0.51     &0.054              \\
\astrobj{Arp\,220}         &Mega              &1.11      &0.32               \\
\astrobj{NML\,Cyg}         &RSG               &-0.33$^@$ &0.041              \\
            \noalign{\smallskip}
            \hline
         \end{tabular}
\begin{list}{}{}
\scriptsize
\item[$\dagger$:] The pumprates are defined as integrated line photon flux ratio 
$n_{\rm OH}/(n_{34.5}+n_{53.3})$ and taken from paper II.
\item[@:] No reliable IRAS fluxes are available to make IRAS color index C$_{32}$, ISO 
               spectra are used to derive equivalent IRAS fluxes.
\end{list}
\end{table}

There are 9 sources in paper I and II showing the 53.3\,$\mu$m absorption line in the ISO LWS spectra but with 
their 34.6\,$\mu$m lines not observed or not detected. They compose the `Sample 2' in this paper. Among them, 
seven ones have 34.6\,$\mu$m ISO SWS spectra but the absorption line is too weak to be detected. In this case, 
3 times of the standard deviation ($3\sigma$ noise) of the continuum can be used to estimate an upper limit 
of the integrated line flux for these missing absorption lines when the line width (FWHM) has been estimated. 
(See how to estimate the line width in spectra of different resolutions in Appendix.) If all these missing 34.6\,$\mu$m 
lines are assumed to be in absorption, no matter how weak they are, the lower limit of the integrated line fluxes should be 
0 W/cm$^2$. The other two sources in sample 2 are not observed by ISO SWS at 34.6\,$\mu$m. In this case, the 
integrated photon flux of the 53.3\,$\mu$m line alone can be considered as a lower limit of the total integrated 
photon flux of both infrared pumping lines, because the missing 34.6\,$\mu$m line can be assumed to be in absorption. 
These upper and lower limits of the integrated infrared absorption line fluxes can then be transformed respectively 
into lower and upper limits of the (pseudo) maser pumprate. The derived pumprates are listed in Table\,\ref{sample2} 
in which column (1) is IRAS name of the sources; column (2) is the source type; column (3) is IRAS color indices 
C$_{32}$, where in the case that no reliable IRAS fluxes are available for defining the color indices ISO spectra 
are used to derive equivalent IRAS fluxes  (See more remarks in the last paragraph 
of this section); column (4) is the integrated 53.3\,$\mu$m line flux (from paper II); column (5) to (7) are 
the ISO SWS spectrum name, standard deviation of 34.6\,$\mu$m continuum and the estimated lower and upper limits 
of the integrated 34.6\,$\mu$m line flux respectively (when more than one ISO SWS spectra are available for a source, 
the one with the lowest $1\sigma$ noise among the brightest ones are used); column (8) is the integrated flux of the OH 
1612\,MHz maser from literature; column (9) is the estimated upper and/or 
lower limit of the pumprate while column (10) gives the reference information and some notes for the OH maser flux data. 
The source types in column (2) are mainly from SIMBAD database while that of the two sources: \astrobj{IRAS 06319+0415} 
and \astrobj{IRAS 17431-2846} are re-identified by this paper (see the details in the next section). Note that, for the red 
supergiant \astrobj{IRAS 03507+1115},  the pumprate is estimated to be in a pair of separated regions: `$> 0.0036$' and 
`$<-0.0008$'. The reason for this strange situation is that the filling emission feature in the 53.3\,$\mu$m line profile of this 
source is so strong that the total flux integrated over the whole line profile becomes net emission! The same curious thing 
also happens for another red supergiant \astrobj{IRAS 07209-2540}, but the difference is that the 34.6\,$\mu$m line 
of this latter source is detected to be very strong absorption, which effectively compensates the net emission in 53.3\,$\mu$m 
line when calculating maser pumprate. How the strong filling emission in the 53.3\,$\mu$m OH absorption line arises is an 
interesting question to answer, but it is not the task of this paper. Eventually, the positive region `$>0.0036$' is chosen for 
\astrobj{IRAS 03507+1115} in further analysis because negative pumprate is impossible for a radiatively pumped OH maser.
%
\begin{table}[]{}
\scriptsize
\caption[]{\scriptsize Sample two: objects with only 53.3\,$\mu$m OH maser pumping line detected. The units of the quantities 
in the table are: [$10^{-19}$W/cm$^2$] for {\it F}$^{int}_{53.3}$ and {\it F}$^{int}_{34.6}$; [$10^{-26}$W/cm$^2$] for 
{\it F}$^{int}_{OH}$; [Jy] for $\sigma_{34.6}$.}
\label{sample2}
\centering
\begin{tabular}{l@{ }l@{  }l@{  }l@{  }c@{  }c@{ }c@{  }c@{  }l@{  }l@{  }}
\hline
\noalign{\smallskip}
Name            &Type$^{\#}$   &C$_{32}$  &{\it F}$^{int}_{53.3}$ &ISO$_{34.6}$ &$\sigma_{34.6}$ &{\it F}$^{int}_{34.6}$ &{\it F}$^{int}_{OH}$ &Pumprate           &Notes$^*$ \\
(1)             &(2)             &(3)       &(4)                    &(5)             &(6)           &(7)                    &(8)                  &(9)                &(10)      \\
\noalign{\smallskip}
\hline
\noalign{\smallskip}
\astrobj{03507$+$1115}      &AGB     &-0.85     &1.37       &80900805    &6.91     &$0\sim-2.55$     &3.24     &$>$0.004/$<$-0.0008    &AOc           \\          
\astrobj{06319$+$0415}      &PPN/PN  &0.41      &-0.83      &$--$        &$--$     &$--$             &0.118    &$<$4.96e-5             &AOa(1p)       \\          
\astrobj{16342$-$3814}      &pAGB    &0.16      &-5.39      &45801328    &5.61     &$0\sim-4.02$     &25.11    &1.10$\sim$1.62E-3      &LC96(fig.)    \\          
\astrobj{17424$-$2852}      &H\,II   &0.99      &-50.34     &$--$        &$--$     &$--$             &1.64     &$<$1.14e-5             &SLW98$^{*1}$  \\          
\astrobj{17430$-$2848}      &Cluster &0.40$^@$  &-21.72     &84101302    &9.48     &$0\sim-3.50 $    &1.81     &2.63$\sim$2.91E-5      &LWH92         \\          
\astrobj{17431$-$2846}      &IR      &0.60$^@$  &-31.6      &67700503    &19.8     &$0\sim-14.17$    &1.86     &1.59$\sim$2.06E-5      &SLW98$^{*2}$  \\          
\astrobj{17574$-$2403}      &H\,II   &0.77      &-93.2      &09901027    &51.9     &$0\sim-43.90$    &41.7     &1.20$\sim$1.56E-4      &BAU79         \\          
\astrobj{18348$-$0526}      &Mira    &-0.14     &-2.82      &47201016    &2.19     &$0\sim-0.68 $    &550      &0.059$\sim$0.068       &AJG74(2p)     \\          
\astrobj{20255$+$3712}      &H\,II   &0.61      &-26.52     &13400330    &22.8     &$0\sim-7.09 $    &0.113    &1.27$\sim$1.49E-6      &AOe(1p)       \\          
            \noalign{\smallskip}
            \hline
         \end{tabular}
\begin{list}{}{}
\scriptsize
\item[Note:] Although \astrobj{IRAS\,06053-0622} was included in paper I and II, it is removed from this paper, 
             because Dr. Peter te Lintel Hekkert recently confirmed (by private communication) that the detection 
             of the 1612MHz maser towards this infrared source is untrustworthy.
\item[*:] Reference codes:                   
          {\bf AJG74}: Andersson et al.\,(\cite{and74});  {\bf AOa}: Chengalur et al.\,(\cite{che93});
          {\bf AOc}: Lewis et al.\,(\cite{lew90});      {\bf AOe}: Lewis\,(\cite{lew94});
          {\bf BAU79}: Baud et al.\,(\cite{bau79a});    {\bf LC96}: te Lintel Hekkert \& Chapman\,(\cite{tel96});
          {\bf LWH92}: Lindqvist et al.\,(\cite{lin92}); {\bf SLW98}: Sjouwerman et al.\,(\cite{sjo98}).
          {\bf (1p)} means only one maser peak was observed; 
          {\bf (2p)} means it is a double peak feature but the literature only give the total integrated maser flux.
          {\bf (fig)} means the integrated 1612\,MHz OH maser fluxes are derived
          by measuring from the figure published in the relevant literature.\\                                 
\item[*1:] For \astrobj{IRAS\,17424-2852}, the closest OH maser used in the paper is \astrobj{OH0.06-0.018}, 
           but the maser position is about 2.34 arcmin away from the SIMBAD position of this infrared source!
\item[*2:] The OH maser \astrobj{OH0.178-0.055} is identified for this IRAS source.
\item[\#:] Most of the object types come from SIMBAD database while that of \astrobj{06319+0415} and 
           \astrobj{17431-2846} are re-identified in this paper (see the detail in the text).
\item[@:] No reliable IRAS fluxes are available to make IRAS color index C$_{32}$, ISO 
               spectra are used to derive equivalent IRAS fluxes.
\end{list}
\end{table}

Many OH/IR sources in paper I and II were observed by both ISO SWS and LWS spectrometers but failed 
to have either of the two maser pumping lines detected. These sources (29 ones) compose the `Sample 3' in this 
paper. The $3\sigma$ noise level of the 34.6 and 53.3\,$\mu$m continuum spectra can supply an estimate 
of the upper limit of the line-center depth of the two absorption lines, if they do exist but are buried by noise. 
Once the width (FWHM) of the two doublets are estimated (see how to estimate them in Appendix), the 
upper limit of the integrated line fluxes can be calculated. These upper limits can again be transformed into lower 
limits of the (pseudo) maser pumprates when the integrated 1612\,MHz maser fluxes are known. The derived 
pumprate lower limits are listed in Table\,\ref{sample3} in which column (1) is IRAS 
name of the sources; column (2) is the source type; column (3) is IRAS color indices C$_{32}$, where in the case 
that no reliable IRAS fluxes are available for defining the color indices, ISO spectra are used to derive 
equivalent IRAS fluxes  (See more remarks in next paragraph); 
column (4) to (6) are the ISO SWS spectrum name, the 34.6\,$\mu$m continuum flux and the relevant standard 
deviation respectively (when more than one ISO SWS spectra are available for a source, the one with the lowest 
$\sigma$ noise among the brightest ones are used); column (7) to (9) are the ISO LWS spectrum name, the 
53.3\,$\mu$m continuum flux and the relevant standard deviation respectively (when more than one ISO LWS 
spectra are available for a source, the one with the spectral resolution power not lower than 1500 and with the best 
quality are used); column (10) is the integrated flux of the OH 1612\,MHz maser from literature; column (11) is the 
estimated lower limit of the pumprate while column (12) gives the 
reference information and some notes for the OH maser flux data.  The source types in column (2) are mainly from 
SIMBAD database while that of 7 sources: \astrobj{10197-5750},  \astrobj{18050-2213}, \astrobj{18498-0017}, 
\astrobj{19343+2926}, \astrobj{20077-0625}, \astrobj{22036+5306} and \astrobj{22177+5936}, are re-identified 
by this paper (see the details below). 

Here I would like to discuss a little more about how the equivalent IRAS fluxes are derived from ISO spectra. Both ISO SWS 
and LWS spectra are composed of several wavelength sub-regions. The last 5 SWS wavelength sub-regions and the first 5 
LWS wavelength sub-regions are needed for deriving IRAS 25 and 60\,$\mu$m photometrical fluxes. Five sources in the paper 
need to determine their IRAS color from ISO spectra (see sources whose C$_{32}$ are marked by `@' in Table\,\ref{sample1}, 
\ref{sample2} and \ref{sample3}). It is found that the SWS and LWS spectra that are taken under different instrumental settings 
agree well with each other in the overlap wavelength region around 45\,$\mu$m for three sources: \astrobj{IRAS\,18498-0017}, 
\astrobj{IRAS\,19039+0809} and \astrobj{NML\,Cyg}. However, LWS spectra are 
found to be several times brighter than SWS spectra in the overlap region for the other two sources: \astrobj{IRAS\,17430-2848} and 
\astrobj{IRAS\,17431-2846}. The large discrepancies are mainly due to three facts: both sources lie in the galactic 
center, the pointing directions of SWS and LWS observations are not strictly the same, and the apertures of the SWS and LWS 
spectrometers are different. The aperture of LWS (diameter of $84''$) is more than 8 times larger than that of SWS ($33''\times 20''$). 
Although I tried to choose SWS and LWS observations pointing as closely as possible, the difference in pointing directions 
between SWS and LWS is still large: $26.''2$ for \astrobj{IRAS\,17431-2846} and $21.''4$ for \astrobj{IRAS\,17430-2848}. 
The galactic center is crowed with all kinds of infrared sources, hence the differences in aperture size and pointing mean that 
the pair of SWS and LWS spectra do not strictly belong to the same source. Therefore, the 
IRAS fluxes derived in such a way for the two galactic center sources should be treated with caution. The IRAS 25 and 60\,$\mu$m 
fluxes derived are: 120 and 299\,Jy for \astrobj{IRAS\,17430-2848}, 361 and 1430\,Jy for \astrobj{IRAS\,17431-2846}, 
88.8 and 187\,Jy for \astrobj{IRAS\,18498-0017}, 190 and 73\,Jy for \astrobj{IRAS\,19039+0809}, 5605 and 2608\,Jy 
for \astrobj{NML\,Cyg}, while the corresponding IRAS color C$_{32}$ (=log({\it F}$_{60}$/{\it F}$_{25}$) ) is 
0.40, 0.60, 0.32, -0.41 and -0.33, respectively.

\section{Source type identification}
\label{identification}

Most of the source types listed in column (2) of Table\,\ref{sample1}, \ref{sample2}, \ref{sample3} are from 
SIMBAD database. But sometimes the source classification in SIMBAD is not clear or not clearly related to the 
evolutionary status of the source (e.g., `variable star', `emission line star', `infrared source', and so on). In order to 
make clear the evolutionary status of the embedded stars associated with these OH masers, identification work is 
performed below on the basis of literature analysis for several ambiguous sources:
%
\begin{table}[]{}
\scriptsize
\caption[]{\scriptsize Sample three: Objects with the non-detection of the maser pumping lines at both 34.6 and 53.3\,$\mu$m. The units 
of the quantities in the table are: [Jy] for {\it F}$^{c}_{34.6}$, {\it F}$^{c}_{53.3}$, $\sigma_{34.6}$ and $\sigma_{53.3}$; 
[$10^{-26}$W/cm$^2$] for {\it F}$^{int}_{OH}$.}
\label{sample3}
\begin{tabular}{l@{ }l@{  }l@{  }l@{  }l@{  }l@{ }l@{  }l@{  }l@{  }l@{  }l@{  }l@{  }}
\hline
\noalign{\smallskip}
Name &Type$^{\#}$ &C$_{32}$ &ISO$_{34.6}$ &{\it F}$^{c}_{34.6}$ &$\sigma_{34.6}$ &ISO$_{53.3}$ &{\it F}$^{c}_{53.3}$ &$\sigma_{53.3}$ &{\it F}$^{int}_{OH}$ &Pumprate &Notes$^*$  \\
(1)  &(2)           &(3)      &(4)          &(5)                  &(6)           &(7)          &(8)                  &(9)           &(10)                &(11)     &(12)       \\
\noalign{\smallskip}
\hline
\noalign{\smallskip}
\centering
\astrobj{01037$+$1219}    &OH/IR  &-0.65       &57700828  &432.6   &1.72  &57701103  &170.4  &2.4   &72.48   &$>$1.56E-2   &AOb                   \\
\astrobj{01304$+$6211}    &Mira   &-0.37       &78800604  &268.2   &3.99  &61300914  &141.8  &3.4   &98.68   &$>$9.63E-3   &B78(fig)              \\
\astrobj{05506$+$2414}    &HH     &0.21        &69901113  &88.2    &1.99  &83901512  &112.3  &2.5   &0.43    &$>$16.86E-5   &AOa                   \\
\astrobj{07027$-$7934}    &PN     &-0.29       &73501035  &92.9    &1.69  &56700908  &29.2   &2.3   &39.01   &$>$8.05E-3   &tL-H(fig)             \\
\astrobj{10197$-$5750}    &PPN/PN &-0.27       &25400259  &470     &5.81  &10300135  &634.7  &5.1   &198.99  &$>$8.74E-3   &SAP93(fig)$^{*1}$     \\
\astrobj{10580$-$1803}    &SRb    &-0.79       &22800918  &130     &1.14  &22800816  &78.7   &4.4   &0.036   &$>$4.99E-6   &ELG03(fig)            \\
\astrobj{15452$-$5459}    &pAGB   &0.05        &45900615  &342.2   &6.04  &48800916  &339.5  &9.3   &89.67   &$>$3.91E-3   &tL-H(fig)             \\
\astrobj{16280$-$4008}    &PN     &0.36        &28901214  &63.7    &1.77  &08402635  &107.2  &4.6   &0.225   &$>$2.92E-5   &LC96(fig)             \\
\astrobj{17103$-$3702}    &PN     &0.4         &28902017  &456     &4.84  &48903704  &846.5  &4.4   &1.88    &$>$2.42E-4   &ZLP89(fig)            \\
\astrobj{17150$-$3224}    &pAGB   &-0.08       &28902214  &408.8   &5.94  &32702239  &294    &5.1   &2.25    &$>$1.13E-4   &ATCAb(fig,1p)$^{*2}$  \\  
\astrobj{17411$-$3154}    &OH/IR  &-0.3        &13601695  &2867.5  &22.7  &46901615  &1382   &5.0   &659.2   &$>$1.76E-2   &GWW89(fig)            \\
\astrobj{17463$-$3700}    &PN     &-0.69       &32400609  &16.6    &2.31  &32400610  &6.4    &4.0   &0.336   &$>$1.96E-5   &SIH95                 \\
\astrobj{18050$-$2213}    &RSG    &-0.72       &09900171  &709.9   &8.33  &33100802  &290.5  &4.4   &430     &$>$2.97E-2   &MBD74(fig)            \\
\astrobj{18196$-$1331}    &YSO    &0.92        &12200841  &1736.7  &11.3  &34000203  &4476   &13.1  &2.65    &$>$9.87E-5   &LC96(fig)$^{*3}$      \\
\astrobj{18272$+$0114}    &OrV*   &-0.47       &51301128  &43.5    &2.04  &14900719  &93.7   &3.1   &0.026   &$>$4.74E-6   &AOe(1p)               \\
\astrobj{18498$-$0017}    &OH/IR  &0.32$^@$  &32001560  &173.1   &3.64  &32300501  &173.7  &4.6   &0.55    &$>$4.76E-5   &AOa                   \\
\astrobj{18560$+$0638}    &Mira   &-0.52       &70900322  &24.9    &4.29  &70900321  &128.7  &3.3   &154.8   &$>$1.47E-2   &AOa                   \\
\astrobj{18596$+$0315}    &OH/IR  &0.2         &32301106  &30      &2.30  &49901207  &32.5   &1.4   &34.86   &$>$5.87E-3   &AOa                   \\
\astrobj{19039$+$0809}    &Mira   &-0.41$^@$   &48903803  &132.8   &1.60  &31600514  &45.6   &1.68  &103.0   &$>$3.44E-2   &SSF88                 \\
\astrobj{19114$+$0002}    &pAGB   &-0.1        &32002241  &765.6   &2.94  &52500806  &660.4  &2.7   &119.3   &$>$2.51E-2   &L89                   \\
\astrobj{19219$+$0947}    &PN     &-0.34       &32002529  &44.7    &1.01  &54700310  &44.7   &2.9   &11.14   &$>$2.33E-3   &ZLP89(fig)$^{*1}$     \\
\astrobj{19255$+$2123}    &PN     &0.22        &17600529  &44.1    &1.96  &17600528  &48.6   &2.4   &2.39    &$>$3.63E-4   &B78(fig)$^{*1}$       \\
\astrobj{19343$+$2926}    &pAGB   &0.3         &52000719  &106.8   &1.73  &36701902  &122.2  &4.6   &0.36    &$>$4.68E-5   &AOa                   \\
\astrobj{20000$+$4954}    &Mira   &-0.8        &37400126  &27.4    &3.13  &26300417  &2.2    &1.8   &2.89    &$>$3.64E-4   &OWMS                  \\
\astrobj{20077$-$0625}    &OH/IR  &-0.69       &17000529  &60.7    &3.56  &34400719  &207.8  &9.1   &15.9    &$>$8.81E-4   &WB72                  \\
\astrobj{22036$+$5306}    &PPN/PN &0.36        &39500297  &90.4    &2.02  &54800798  &119    &6.2   &4.06    &$>$3.34E-4   &LC96(fig)             \\
\astrobj{22176$+$6303}    &H\,II  &0.86        &06401146  &2738.2  &16.6  &82301123  &11792  &39.5  &0.16    &$>$2.39E-6   &BLS90(fig,1p)         \\
\astrobj{22177$+$5936}    &OH/IR  &-0.4        &28300921  &158     &2.90  &28300920  &94.2   &3.0   &134.2   &$>$1.50E-2   &BHMW                  \\
\astrobj{23412$-$1533}    &Mira   &-0.91       &18100530  &202.8   &3.02  &20000913  &67     &2.74  &0.04    &$>$4.99E-6   &ISH94                 \\
            \noalign{\smallskip}
            \hline
         \end{tabular}
\begin{list}{}{}
\scriptsize
\item[*:] Reference codes: AOa, AOe, LC96 are the same as in Table\,\ref{sample2} while the others are: 
		{\bf AOb}: Eder et al.\,(\cite{ede88}); {\bf ATCAb}: Sevenster et al.\,(\cite{sev97});
          {\bf BHMW}: Baud et al.\,(\cite{bau79b}); {\bf BLS90}: Braz et al.\,(\cite{bra90});
          {\bf B78}: Bowers\,(\cite{bow78}); {\bf ELG03}: Etoka et al.\,(\cite{eto03});
          {\bf GWW89}: Gaylard et al.\,(\cite{gay89}); {\bf ISH94}: Ivison et al.\,(\cite{ivi94});
          {\bf L89}: Likkel\,(\cite{lik89}); {\bf MBD74}: Masheder et al.\,(\cite{mas74});
          {\bf OWMS}: Olnon et al.\,(\cite{oln80}); {\bf SAP93}: Silva et al.\,(\cite{sil93});
          {\bf SIH95}: Seaquist et al.\,(\cite{sea95}); {\bf SSF88}: Sivagnanam et al.\,(\cite{siv88});
          {\bf tL-H}: te Lintel Hekkert et al.\,(\cite{tel91}); {\bf WB72}: Wilson \& Barrett\,(\cite{wil72});
          {\bf ZLP89}: Zijlstra et al.\,(\cite{zij89}).
          {\bf (1p)} means only one maser peak was observed; 
          {\bf (fig)} means the integrated 1612\,MHz OH maser fluxes are derived
          by measuring from the figure published in the relevant literature.\\                                 
\item[*1:] No data for the red maser peak, so only the blue peak is used.
\item[*2:] The maser \astrobj{OH353.844+02.984} is identified for \astrobj{17150-3224}.
\item[*3:] For \astrobj{18196-1331}, the OH 1612\,MHz maser show four peaks, here the sum of integrated maser fluxes of 
           the two low velocity peaks and the sum of the two high velocity peaks are presented in column (10).
\item[\#:] Most of the object types come from SIMBAD database while that of \astrobj{10197-5750}, 
             \astrobj{18050-2213}, \astrobj{18498-0017}, \astrobj{20077-0625}, 
             \astrobj{22036+5306} and \astrobj{22177+5936} are re-identified in 
             this paper (see the detail in the text).
\item[@:] No reliable IRAS fluxes are available to make IRAS color index C$_{32}$, ISO 
               spectra are used to derive equivalent IRAS fluxes.
\end{list}
\end{table}

{\it \astrobj{IRAS 06319+0415}}: SIMBAD assigned this source to be an unknown infrared source. 
Lada and Gautier\,(\cite{lad82}) studied the energetic molecular outflow of this source by millimeter-wave and 
infrared observations. They found that the source, which is located in the Rosette molecular cloud outside the 
solar circle, shows anisotropic high velocity outflow and radio continuum. They reviewed that this source, with 
no visible counterpart, might be a nonstellar object. But the very strong Br$\alpha$ recombination line indicates that 
it is not a usual H\,II region. Lenzen et al.\,(\cite{len84}) performed high resolution infrared imaging 
to this source and found that it is a double source with the main direction of its CO-mass outflow nearly perpendicular 
to the direction of the separation of the two components. Castelaz et al.\,(\cite{cas85}) also found from near infrared imaging 
that this source is composed of two parts: a compact H\,II region and a very young prestellar object. They suggested 
that it is very likely a binary system.  Colin\,(\cite{col98}) further pointed out from the image of multiple non-axially 
symmetric near-IR bow shocks around this source that both components of the binary system can produce actively 
driven outflow in different phases. Phelps and Lada\,(\cite{phe97}) obtained J, H, K near infrared image for the source 
with higher spatial resolution and  identified it to be a cluster embedded in the Rosette Molecular Cloud. From 
these pre-existed studies, it can be concluded that this source is a peculiar H\,II region that shows active outflow and very 
possibly resides in a binary system or even in cluster.

{\it \astrobj{IRAS 10197-5750}}:  This source is classified as Wolf-Rayet (WR) star in SIMBAD, but the 
classification is wrong. Roberts\,(\cite{rob62}), for the purpose of statistical work, tentatively classified it as 
a WR star on the base of unpublished finding of widened H$\alpha$ emission, and thereby it gained the popular 
name Roberts 22 from his catalogue. However, the signpost of WR star should be Helium emission line instead of 
H$\alpha$. 
Recent works such as that of Sahai et al.\,(\cite{sah99}) and Zijlstra et al.\,(\cite{zij01}) studied the morphology 
of this source by high resolution imaging with the Hubble Space Telescope and the Australia Telescope Compact 
Array and proved that it is a PPN with bipolar outflow and with OH maser points located on the edge of a torus-like 
disk. Dyer et al.\,(\cite{dye01}) reviewed the physical status of this source and concluded from the already known 
optical, OH maser and near infrared observations, and especially from the presence of Fe\,II absorption, the absence of 
He\,I and the weakness of the Balmer absorption lines that it might be a PPN with a hidden A2 Ie central star. They also 
demonstrated in an astrometric way that the OH masers arise in the equatorial waist of the optical bipolar outflow and the maser 
kinematics support the classification of PPN. Garcia-Hernandez et al.\,(\cite{gar02}) also demonstrated from the 
H$_2$ emission from this source that it is a post AGB star. From these previous works, this source should be 
classified as an OH-PPN.

{\it \astrobj{IRAS 17431-2846}}: This source is close to Galactic Center (GC). Many infrared sources crowd around 
the source position and both the IRAS 
and ISO SWS/LWS beams are large enough to include more than one IR sources. So it is difficult to associate 
the observed IRAS or ISO data to a single star. Therefore, in this paper, this source keeps the original classification 
from SIMBAD: infrared source. But its OH 1612\,MHz maser is considered as interstellar OH maser in this paper, 
because it  is very close to GC where a lot of interstellar OH masers exist.

{\it \astrobj{IRAS 18050-2213}}: SIMBAD classifies this source as a Semi-Regular (SR) variable.  Lockwood 
\& Wing\,(\cite{loc82}) concluded from the multi-wavelength light variation and spectral type of this source that it is 
a cool supergiant. Further more, SiO and H$_2$O masers have been found by Yu\,(\cite{yu92}), 
Greenhill et al.\,(\cite{gre95}) and Pashchenko \& Rudnitskii\,(\cite{pas99}), which further confirms its late 
type star nature.

{\it \astrobj{IRAS 18498-0017}}: SIMBAD classifies this source as a Mira variable.  Nyman et al.\,(\cite{nym93}) 
did not detect SiO maser towards this source. But Palagi et al.\,(\cite{pal93})  did find H$_2$O maser towards it. 
Groenewegen\,(\cite{gro94}) used the distance of 4.8 kpc to constrain a luminosity of 15800 L$_{\odot} $ for it. 
Combining with a very long light variation period of 1539 days and a huge mass loss of $1.6\times 10^{-4}$ M$_{\odot}$/yr, 
Groenewegen identified this source to be an extreme OH/IR star. This identification also agrees to the lack of SiO 
maser and the finding of H$_2$O maser, that is to say, it may be on the way  leaving AGB.

{\it \astrobj{IRAS 20077-0625}}: SIMBAD classifies this source as a Mira variable.  Hashimoto \& Izumiura\,(\cite{has97}) 
reported that this source looks like a compact stellar source on the 60 and 90\,$\mu$m images. They also judged from 
the shape of the spectral energy distribution (SED) and very large mass loss rate ($1.4\times 10^{-5}$M$_{\odot}$/yr) 
that it is an OH/IR star on the top of AGB.

{\it \astrobj{IRAS 22036+5306}}: SIMBAD classifies this source as an OH/IR star.  However, 
Sahai et al.\,(\cite{sah03}) argued that this is a PPN because they found that the central star belongs to F5 spectral 
type and shows optical H$_{\alpha}$ and H$_{\beta}$ emission and there exists an extended bipolar nebula around 
it. Zijlstra et al.\,(\cite{zij01}) confirmed that the OH maser occurs in a shell and a linear outflow and took it as a 
typical OH-PN. Therefore it should be an OH-PPN/PN rather than an OH/IR star.

{\it \astrobj{IRAS 22177+5936}}: SIMBAD classifies this source as a variable star. Sylvester et al.\,(\cite{syl99})  
observed the circumstellar dust radiation of this source by ISO and determined a mass loss rate of 
$2.4\times 10^{-7}$M$_{\odot}$/yr together with a color temperature of 400K for the circumstellar envelope. 
The mass loss rate is typical for a Mira or OH/IR star of small mass. Maser observations resulted in the detections 
of SiO (Cernicharo et al.\,\cite{cer97}) and H$_2$O (Gonz$\acute a$lez-Alfonso et al.\,\cite{gon98}) masers 
towards this source. Van Langevelde et al.\,(\cite{van90}) derived a luminosity of $1.17\times 10^{4}$L$_{\odot}$ 
using the distance determined by OH maser phase lag method and an infrared variation period of 1460 day for the 
source, which demonstrated that the star can not be a Mira or a red supergiant but a small mass OH/IR star on or 
near the top of AGB. The fact that Jiang et al.\,(\cite{jia97}) tried but failed to identified the optical counterparts 
of the star up to 20 mag in V band and 19 mag in I bands also supports that it is a star with very thick circumstellar 
envelope: OH/IR star.

\section{The OH maser pumprate of different types of source}
\label{pumprate}

\begin{figure}[]
   \centering
	 \includegraphics[width=9cm]{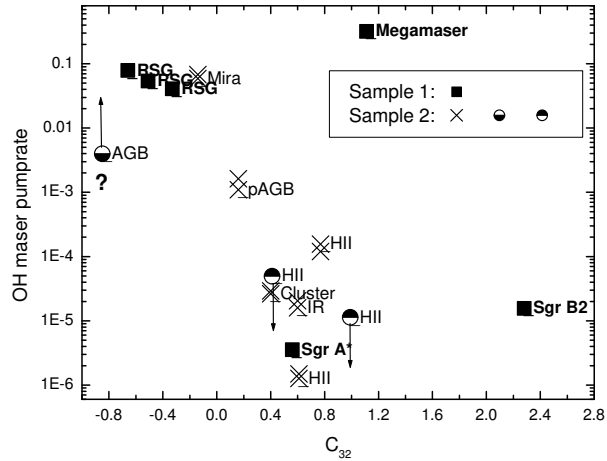}
   \caption{OH 1612\,MHz maser pumprate versus IRAS color C$_{32}$ for sample 1 and 2 sources. Source types are marked beside each source. Double crosses represent the pumprate ranges determined for some sample 2 sources while arrows mean that the pumprate value is merely an upper or lower limit. The question mark cautions the readers that only the positive lower limit of the maser pumprate for this source is shown in the figure while another negative upper limit (see Table\,\ref{sample2}) is dropped due to the unreality of the negative pumprate for a radiatively pumped OH maser.}	
   \label{figsample12}
   \end{figure}

One of the main tasks of this paper is to investigate whether the late type stars whose OH 1612\,MHz masers are thought to be radiatively pumped show proportional relationship between the IR absorption and the maser emission (i.e., show fixed pumprate) and whether the H\,IIs whose OH 1612\,MHz maser can be pumped by variety of  mechanisms violate this relationship (i.e., show varying pumprates). The pumprates of sample 1 sources and the pumprate ranges of sample 2 sources are shown in Fig.\,\ref{figsample12}. The distribution of the six sample 1 sources can be grouped obviously into 3 regions: the three RSGs lie in the upper left region, the two galactic center (GC) sources lie in the lower left region while the only megamaser \astrobj{Arp\,220} lies high up in the middle top of the figure. The large differences in the (pseudo) radiative pumprates of different types of maser sources had already been addressed in paper I and II. Taking account the sample 2 sources, one can see in Fig.\,\ref{figsample12} that the separation of stellar OH masers (RSGs, Mira) from interstellar OH masers (H\,IIs, cluster, GC sources) still holds while two post-AGB sources (pAGB, PPN) lie between the two regions. One should note that the three RSGs and the Mira crowd tightly around the pumprate of about 0.06 while the interstellar OH masers distribute in a large area with their (pseudo) OH maser pumprates varying for almost two orders of magnitude. 

\begin{figure}[]
   \centering
	 \includegraphics[width=9cm]{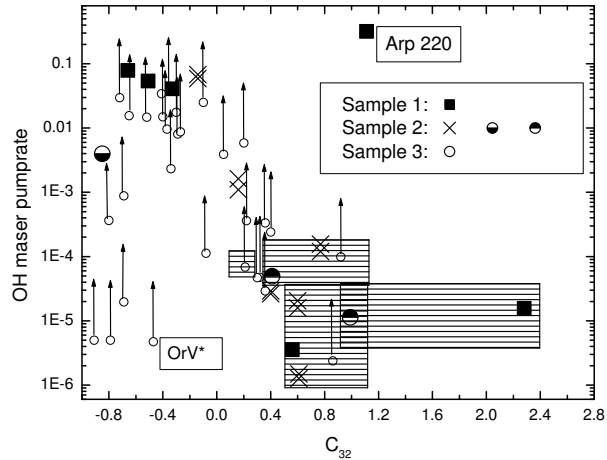}
   \caption{OH 1612\,MHz maser pumprate versus IRAS color C$_{32}$ for all sample 1, 2 and 3 sources. The arrows mean that the pumprates for all sample 3 sources are lower limits only. The hatched square areas represent the distribution regions of interstellar OH maser sources while the other sources, with the exception of the two ones marked by their type or name: OrV* and \astrobj{Arp\,220}, all belong to stellar OH masers.}
   \label{figsample123}
   \end{figure}

\begin{figure}[]
   \centering
	 \includegraphics[width=9cm]{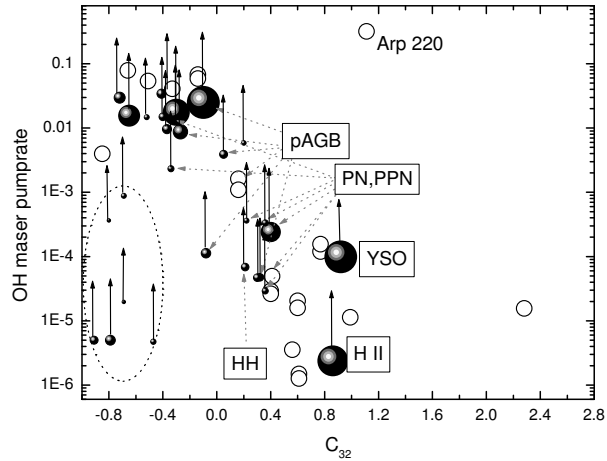}
   \caption{The same as Fig.\,\ref{figsample123} but with different marks: all sources of sample 1 and 2 are shown in open circles while all sample 3 sources are shown in black balls with the ball size proportional to the mean S/N ratio of the infrared spectra. The source types of some sample 3 sources that have high S/N ratio (large size balls) are marked in the figure. Those marked by gray dotted-line arrows are post AGB stars or PPNe/PNe which mainly distribute in the transitional region between the stellar OH masers in the upper left region and the interstellar OH masers in the lower right region.}	
   \label{figlow}
   \end{figure}

This difference in variation range of the OH maser pumprates sheds light on the maser pumping mechanisms of the sources. The nearly fixed pumprates of stellar OH 1612\,MHz masers supports the argument that they are really mainly radiatively pumped, because the absorption of IR photons is proportional to the emission of maser photons. But the low value and large variation of maser pumprates of interstellar masers may be due to several possible reasons. Firstly, the most significant factor is the difference in physical environments. Compared with late type stars, SFRs usually have more extended, thicker and colder circumstellar envelope. The OH molecules can be produced not only in the inner edge of the dusty envelope by either high temperature stellar radiation or by shock waves, but also in the outer edge of the envelope by interstellar UV radiation (similar to that occurs around late type star). Maybe not all the OH molecular cloudlets in SFRs can successfully stimulate maser action. For example, Szymczak \& G$\acute e$rard\,(\cite{szy04a}) found 36 H\,II regions to show blue shifted OH main line absorption, and argued that these absorptions may be caused by the OH molecules formed in front of the central continuum sources. Lets call such non-maser OH cloudlets {\it quiet OH cloudlets}. Therefore, if a significant number of OH cloudlets in H\,IIs are quiet, the pseudo pump rate will become small and change with changing number of  quiet OH cloudlets. Secondly, the difference in maser pumping mechanisms can also contribute to the large variation of the (pseudo) maser pumprates. The OH maser pumping mechanism in H\,IIs was incipiently considered collision because the masing OH molecules were believed to be produced in the shock compressed geometrically thin layer between the inner ionized hydrogen sphere and the outer dusty envelope (Elitzur\,\cite{eli92}). However, later works (e.g., Pavlakis \& Kylafis\,\cite{pav96a}, \cite{pav96b}) showed that the radiative OH maser pumping involving local and non-local line overlap effects are also possible for H\,IIs, because there usually exists a remarkable velocity field in the OH masing region therein. Further more, Thissen et al.\,(\cite{thi99}) and Liu et al.\,(\cite{liu04}) demonstrated by their dynamical calculations that the chemical pumping of OH masers is also possible in H\,IIs. It is evident that the quite different OH maser pumping mechanisms will involve quite different energy level transition schemes among the OH rotational levels, which may result in quite different (pseudo) maser pumprates, so as to explain the scattered distribution of interstellar OH masers in Fig.\,\ref{figsample12}, \ref{figsample123} and \ref{figlow}. Thirdly, many interstellar OH masers are found to be strong mainline OH masers (e.g., Caswell \& Haynes\,\cite{cas83}). Because the 1665 and 1667\,MHz mainline maser transitions share the upper and lower energy levels with the 1612\,MHz transition respectively, the strong mainline masers may weaken the 1612\,MHz maser through the so-called competitive gain (Field\,\cite{fie85}), and hence result in very small 1612\,MHz pump rate. Lastly, the anisotropy of the maser region may also play a role in the small and varying pump rates. Most interstellar OH masers are found to show irregular morphology, and sometimes bipolar morphology (e.g., Hutawarakorn et al.\,\cite{hut02}).  OH maser is usually strongly beamed. If the strongest beam is not pointing to us, the observed maser flux can significantly decrease, and then results in very small pump rates and the pump rates will highly vary with varying maser beam pointing.

One may have noted that a post AGB star and a PPN in Fig.\,\ref{figsample12} lie between the two groups of masers: stellar OH masers and interstellar OH masers. This hints that the pumprates of these post AGB objects may form a bridge connecting the two typical kinds of OH masers due to some unknown reasons. This point will be confirmed below when the sample 3 sources are included in Fig.\,\ref{figsample123} and \ref{figlow}.

The sample 3 sources only have lower limit estimation of the (pseudo) OH maser pumprate. These lower limits are compared with the more accurate pumprates or pumprate ranges of the sources from sample 1 and  2 in Fig.\,\ref{figsample123}. The distribution areas of interstellar OH masers, including those arising in H\,IIs, SFRs, Young Stellar Objects (YSOs), an Herbig-Haro object (HH), a stellar cluster near GC, an unknown infrared source near GC and the two GC sources, are shown with hatched squares in Fig.\,\ref{figsample123}. One can see from the figure that, on the one hand, the interstellar OH masers are well separated from the stellar OH masers and show very small (pseudo) maser pumprates, on the other hand, these lower limits seem also to follow the decreasing trend from stellar OH masers in the upper left region to the interstellar OH masers in the lower right region. However, a few sources in the lower left corner have the typical IRAS C$_{32}$ color of stellar OH maser but show abnormally small pumprate lower limits. 

The behavior of the maser pumprate lower limit distribution in Fig.\,\ref{figsample123} is not only related to the nature of the different types of OH masers but also largely affected by the discrepancy between the maser pumprate lower limits and the true maser pumprates. However, this unknown discrepancy can be partly reflected by the signal to noise (S/N) ratio of the infrared spectra. The S/N ratio is defined for the 34.6 and 53.3\,$\mu$m continuum fluxes respectively using the $1\sigma$ noise listed in Table\,\ref{sample3}, and then a mean S/N ratio is defined as the mathematical average of  the two. The S/N ratios of sample 3 sources turn out to be varying quite a lot (ranging from 4.4 to 252.5). One can find from Table\,\ref{sample3} that the $1\sigma$ noise at the two wavelengths do not vary too much. The large variation of the S/N ratios mainly originates from the variation of the IR brightness of the sources. Therefore, the sources with lower S/N ratios are mainly those with lower IR brightness and hence with smaller OH absorption depths (if all sources are assumed to have approximately the same {\it relative} OH absorption depth). In such sense, the maser pumprate lower limits derived from spectra with higher S/N ratio are generally closer to the true maser pumprates while those derived from spectra with very poor S/N ratio is generally far smaller than the true maser pumprates. 

To visualize the effect of the S/N ratios, Fig.\,\ref{figsample123} is re-drawn in Fig.\,\ref{figlow} in which the sample 3 sources are shown in black balls with the ball size proportional to the S/N ratio while the sample 1 and 2 sources are all shown in open circles. One can easily see from this figure that the positions of the largest balls, i.e., those sample 3 sources with the maser pumprate lower limits most close to the true maser pumprates, are well coincident with the distribution regions of stellar and interstellar OH masers from sample 1 and 2, which confirms once again the difference in OH maser pumprates between the two typical kinds of OH masers. Those stellar-OH-maser mimics with unexpected too small maser pumprate lower limits (i.e., sources located in the dotted line circle in Fig.\,\ref{figlow}) are all sources with poor S/N ratios (with small ball sizes in the figure). That is to say, the too small maser pumprate lower limits of these sources should be blamed to the too weak apparent IR brightness and hence the too poor quality of the spectral data.

In Fig. 3, all post-AGB stars and PNe/PPNe are marked by gray dotted-line arrow. As one can see, these post AGB sources mainly occupy the region between the two regions of stellar and interstellar OH masers. This fact confirms the earlier guess that the post AGB sources may have intermediate pumprates to bridge the two typical kinds of OH masers. This transitional behavior of maser pumprate may be due to the physical changes happened in post AGB stars, e.g., the change of the envelope morphology, the detachment of the inner edge of the circumstellar envelope, which results in the sharp decrease of the amount of hot dust grains that are responsible to emit most of the infrared maser-pumping photons, and so on. However, this bridging phenomenon is mainly supported by the lower limit estimations of the maser pumprates of sample 3 sources, it is necessary to obtain more accurate OH maser pumprates for these sources in the future to further confirm this finding.

\section{Discussion}
\label{discussion}

OH 1612\,MHz maser pump rates of stellar masers are found to be nearly constantly around 0.06 from this paper. The model of Elitzur et al.\,(\cite{eli76}) predicted that five to ten 34.6\,$\mu$m photons are needed to produce a maser photon (corresponding to a pump rate of $0.1\sim 0.2$). Realizing that the pump rates derived in this paper have taken account the contribution of 53.3\,$\mu$m line, the derived pump rates basically agree with the theoretical prediction. If all energy level transition processes, such as collision, local and non-local line overlap, are taken account in the theoretical model and the maser pump rate can be defined in the same way as in this paper, it is reasonable to expect a better agreement between the theoretical and observed OH maser pump rates for stellar masers.

Previous statistical works usually defined the 1612\,MHz maser pump rate using the 35\,$\mu$m photometrical data and the peak maser flux intensity. This definition is only a loose lower limit of the true pump rates, hence the derived `pump rates' strongly varied from star to star. For example, both Silva et al.\,(\cite{sil93}) and Chen et al.\,(\cite{che01}) used the definition $S_{OH}/S_{35}$. The former derived pump rates ranging from $0.03\sim 0.25$ for a sample of OH/IR stars with very cool envelope, while the latter found that the pump rates were increasing with IRAS color  log($F_{25}/F_{12}$) and distributed in a broad range from 0.001 to 1 for a sample of OH/IR stars with type E or type A IRAS/LRS spectrum. Werner et al.\,(\cite{wer80}) used the same definition and found a pump rate range $0.16\sim 0.4$ for five OH/IR stars with very thick envelope. The large variations among these `pump rates' are mainly due to the fact that the `pump rate' is only a coarse lower limit. One may note that the pump rates found by these authors are usually far larger than the value 0.06 found in this paper. This is because the pump rate is defined differently in this paper using integrated photon fluxes. Some authors also tried to defined pump rate with integrated line fluxes in an approximate way: $\alpha=F_{OH}\Delta V_{OH}/F_{35}V_e$. But the pump rate defined in this way also suffers large uncertainty and varies strongly. For example, Le Bertre et al.\,(\cite{leb84}) derived such pump rates of 0.05 for three type\,II OH/IR stars while David et al.\,(\cite{dav93}) found in similar way that the maser pump rates range from 0.0005 to 0.25 (note that these pump rates are defined by energy fluxes instead of photon fluxes). In this paper, the properly defined 1612\,MHz maser pump rates shown for the four stellar OH masers in Fig.\,\ref{figsample12} distribute in a narrow range ($0.041\sim 0.079$), which forcefully confirms the radiative pumping nature of these stellar OH 1612\,MHz masers. However, the lower limits of the pump rates of stellar OH masers (sources with small color indices in Fig.\,\ref{figsample123} and \ref{figlow}) still distribute in a broad range ($10^{-5}\sim 0.1$), which demonstrates that the lower limit of OH 1612\,MHz maser pump rate estimated using spectral noise level is already not reliable, not to mention those defined by photometrical data.

The fact that interstellar OH masers show smaller 1612\,MHz maser pump rates had been addressed by other authors using the pump rate defined by IR photometrical data. Le Bertre et al.\,(\cite{leb84})  found that the maser pump rate defined as $\alpha=F_{OH}\Delta V_{OH}/F_{35}V_e$ is larger than 0.01 for three type II OH/IR stars but smaller than 0.01 for a type I source. Dickinson\,(\cite{dic87}) used the definition $S_{OH}/S_{35}$ and found that the pump rates range from about 0.07 to 0.2 for type II OH/IR stars but are about 0.02 for type I sources. However, these contrast between type I and type II OH/IR sources are not as prominent as that found in this paper. From Fig.\,\ref{figsample12}, \ref{figsample123} and \ref{figlow}, one can see that the properly defined 1612\,MHz maser pump rates vary across 5 orders of magnitude from stellar OH masers (type II OH/IR sources) to interstellar OH masers (type I OH/IR sources). As discussed in previous section, one of the possible explanations for the very low maser pump rates of interstellar OH masers is the so-called competitive gain between mainline and 1612\,MHz OH masers. Here I would like to supply more observational proofs upon this point by comparing the peak strengths of the two competing kinds of masers.

\begin{figure}[]
   \centering
	 \includegraphics[width=9cm]{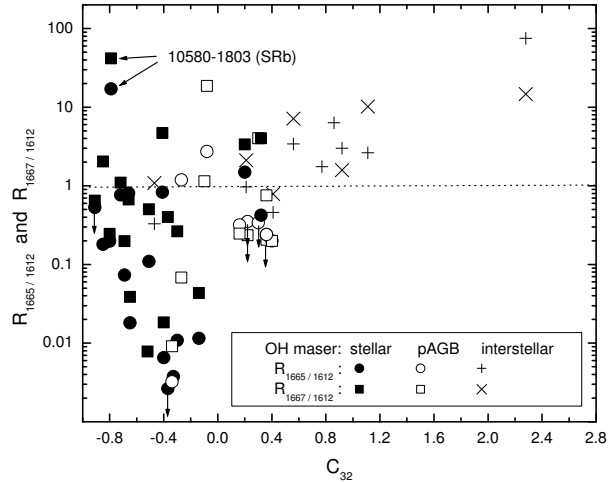}
   \caption{Comparison between 1612\,MHz and mainline OH masers shown against IRAS color. $R_{1665/1662}$ and $R_{1667/1612}$ are the ratio of the maser peak fluxes. The dotted line marks the border above which mainline maser is stronger than 1612\,MHz maser. Note that pairs of marks with the same color belong to one and the same source. Arrows mean only upper limits of mainline maser emission is provided.}
   \label{figmainline}
   \end{figure}

The interstellar OH masers are usually found to show stronger mainline OH masers than 1612\,MHz maser. The 1665 and 1667\,MHz mainline maser transitions share upper and lower energy levels with 1612\,MHz transition respectively. Therefore, in the case that mainline OH masers are as strong as or even stronger than the 1612\,MHz maser, the latter should has been significantly weakened by the former through competitive gain (Field\,\cite{fie85}), which eventually results in very small 1612\,MHz pump rate. In order to verify that the interstellar OH masers in sample 1, 2 and 3 are really strong mainline masers, the peak intensities of strongest peak of mainline masers and low velocity peak of 1612\,MHz masers are collected for most sample sources from literature (see in Table\,\ref{mainline} in which the literature for the 1612\,MHz peak flux data are the same as those from which the integrated maser fluxes are extracted). The mainline to 1612\,MHz peak flux ratios are visualized in Fig.\,\ref{figmainline} in which the sources in different evolutionary status are differentiated by different symbols. As one can see, most stellar OH masers (filled circles and squares) have 1612\,MHz maser stronger than mainline masers while it is reversed for interstellar masers (pluses and crosses). This confirms that the interstellar OH masers in the sample are really strong mainline emitters. Therefore, the competitive gain between mainline and 1612\,MHz masers should be a rather reasonable explanation for the very small 1612\,MHz pump rates of interstellar OH masers, although it is still not clear how the pumping process of the mainline masers affects the 34.6 and 53.3\,$\mu$m absorption lines. A special sources \astrobj{IRAS\,10580-1803} should be noted. Although it is classified as stellar OH maser, it show much stronger mainline OH maser in Fig.\,\ref{figmainline}. It is a SRb variable. The circumstellar envelope of SRb is believed to be even optically thinner than that of Mira variables, and hence it usually emits stronger mainline masers. 

The last point to note is the role of filling emission that is found in 53.3\,$\mu$m absorption line profiles of both stellar and interstellar OH masers (He \& Chen\,\cite{he04b}). The 1612\,MHz maser pump rates calculated in the paper have taken account the contribution of the filling emission feature. For two stellar masers, \astrobj{IRAS\,03507+1115} and \astrobj{IRAS\,07209-2540}, the filling emission is so strong that the 53.3\,$\mu$m line becomes net emission. For the latter source, the very strong absorption at 34.6\,$\mu$m balanced the emission at 53.3\,$\mu$m and a 1612\,MHz maser pump rate typical of stellar OH maser is found (0.079). However, for the former source, the estimated upper limit of the pump rate becomes negative due to the non-detection of the 34.6\,$\mu$m absorption (the $3\sigma$ noise of the 34.6\,$\mu$m continuum spectral flux is low). It seems that some unknown processes are at work not only in the pumping of the observed 1612\,MHz maser but also in producing the strong filling emission in 53.3\,$\mu$m OH line. It is worthy of further maser pumping modeling work.
\begin{table}[]{}
\scriptsize
\caption[]{\scriptsize OH maser peak fluxes of the sample sources (from literature).}
\label{mainline}
\centering
\begin{tabular}{l@{ }l@{  }l@{  }l@{  }l@{  }|l@{ }l@{  }l@{  }l@{  }l@{  }}
\hline
\noalign{\smallskip}
Name &$F^{p}_{1612}$ &$F^{p}_{1665}$ &$F^{p}_{1667}$ & Ref &Name &$F^{p}_{1612}$ &$F^{p}_{1665}$ &$F^{p}_{1667}$ & Ref \\
     &Jy             &Jy             &Jy             &     &     &Jy             &Jy             &Jy             &     \\
\noalign{\smallskip}
\hline
\noalign{\smallskip}
\multicolumn{4}{l}{\bf Sample\,1}                 &          &\astrobj{15452$-$5459} &13.99 &         &         &          \\
\astrobj{07209$-$2540} &200   &162      &134      &BM82(fig) &\astrobj{16280$-$4008} &0.29  &$<$0.07  &$<$0.06  &LC96      \\
\astrobj{17424$-$2859} &0.36  &1.23     &2.57     &LC96      &\astrobj{17103$-$3702} &0.55  &0.11     &0.11     &PPT       \\
\astrobj{17441$-$2822} &1.7   &128      &25       &CH83*1    &\astrobj{17150$-$3224} &1.35  &3.7      &25.2     &HTS       \\
\astrobj{19244$+$1115} &45.6  &5        &23       &NB92(fig) &\astrobj{17411$-$3154} &134.9 &1.47     &35.54    &ELG04     \\
\astrobj{Arp\,220}     &0.032 &0.084    &0.33     &BH87      &\astrobj{17463$-$3700} &int*3 &Non      &         &LDL       \\
\astrobj{NML\,Cyg}     &638   &2.4      &         &ED04      &\astrobj{18050$-$2213} &1.82  &1.4      &2        &CC86      \\
\multicolumn{4}{l}{\bf Sample\,2}                 &          &\astrobj{18196$-$1331} &0.6   &1.8      &0.95     &LC96      \\
\astrobj{03507$+$1115} &2.29  &0.41     &4.69     &L97       &\astrobj{18272$+$0114} &0.068 &0.022    &0.074    &AOe       \\
\astrobj{06319$+$0415} &0.082 &0.038    &0.065    &L97       &\astrobj{18498$-$0017} &0.38  &0.16     &1.52     &DT91      \\
\astrobj{16342$-$3814} &9     &2.86     &2.22     &LC96      &\astrobj{18560$+$0638} &70    &         &0.55     &SHH       \\
\astrobj{17424$-$2852} &0.69  &         &         &          &\astrobj{18596$+$0315} &0.37  &0.54     &1.23     &L97       \\
\astrobj{17430$-$2848} &0.96  &         &         &          &\astrobj{19039$+$0809} &7.82  &6.5      &36.69    &EL00      \\
\astrobj{17431$-$2846} &0.66  &         &         &          &\astrobj{19114$+$0002} &43.8  &         &50       &SG04      \\
\astrobj{17574$-$2403} &34.7  &60.85    &         &FC99*2    &\astrobj{19219$+$0947} &5     &0.016    &0.046    &AOe       \\
\astrobj{18348$-$0526} &400   &4.6      &17.38    &ELG04     &\astrobj{19255$+$2123} &1.7   &$<$0.6   &0.4      &ESWW      \\
\astrobj{20255$+$3712} &0.19  &-1.15    &-1.66    &AOe       &\astrobj{19343$+$2926} &0.29  &$<$0.1   &1.18     &STZW      \\
\multicolumn{4}{l}{\bf Sample\,3}                 &          &\astrobj{20000$+$4954} &2.37  &0.47     &0.58     &OWMS(fig) \\
\astrobj{01037$+$1219} &60.4  &1.09     &2.35     &L97       &\astrobj{20077$-$0625} &35.3  &2.6      &7        &OWMS(fig) \\
\astrobj{01304$+$6211} &32.1  &$<$0.085 &12.9     &L89       &\astrobj{22036$+$5306} &0.83  &0.2      &0.63     &LC96      \\
\astrobj{05506$+$2414} &0.18  &0.17     &0.38     &L97       &\astrobj{22176$+$6303} &0.34  &2.16     &Non      &BLS90     \\
\astrobj{07027$-$7934} &8.62  &         &         &          &\astrobj{22177$+$5936} &41.3  &0.27     &0.76     &ELG04     \\
\astrobj{10197$-$5750} &42.2  &50.1     &2.88     &DGK       &\astrobj{23412$-$1533} &0.06  &$<$0.032 &$<$0.039 &NHWA      \\
\astrobj{10580$-$1803} &0.18  &3.12     &7.62     &ELG03     &                       &      &         &         &          \\
            \noalign{\smallskip}
            \hline
         \end{tabular}
\begin{list}{}{}
\scriptsize
\item[Note:] Blank means no data available. `Non' means non-detection and no 
             upper limit is provided by literature. 
             Upper limits are $1\sigma$ noise for non-detections.
\item[*1:] The mainline OH maser \astrobj{OH\,0.66-0.04} is identified for \astrobj{IRAS\,17441-2822}
\item[*2:] Totally 53 OH 1665\,MHz maser spots are found in the H\,II region \astrobj{IRAS\,17574-2403} by 
  Forster and Caswell\,(\cite{for99}). The beam size of the 1612\,MHz maser observation by 
  Baud et al.\,(\cite{bau79a}) is much larger than that of the mainline maser observations. The 1612\,MHz maser 
  should correspond to the whole 53 mainline maser spots. Therefore the sum of the 53 mainline maser peak fluxes 
  is presented in the table.
\item[*3:] Only integrated mainline maser flux is available and no maser profile figure was published.
\item[(fig):] The flux is derived by measuring directly from the maser line profile figure in literature.
\item[Reference codes:] (The literature are for mainline data. The 1612\,MHz data are for low 
      velocity peak and come from the same literature from which the integrated fluxes are 
      extracted.)
{\bf BH87: }Baan et al.\,(\cite{baa87});         
{\bf BLS90:} Braz et al.\,(\cite{bra90});        
{\bf BM82: }Benson et al.\,(\cite{ben82});       
{\bf CC86: }Chapman et al.\,(\cite{cha86});      
{\bf CH83: }Caswell et al.\,(\cite{cas83});      
{\bf DGK:  }Dyer et al.\,(\cite{dye01});          
{\bf DT91: }Dickinson et al.\,(\cite{dic91});    
{\bf ED04: }Etoka \& Diamond\,(\cite{eto04a});   
{\bf EL00: }Etoka et al.\,(\cite{eto00});        
{\bf ELG04:} Etoka et al.\,(\cite{eto04b});      
{\bf ELG03:} Etoka et al.\,(\cite{eto03});       
{\bf EN79: }Engels D.\,(\cite{eng79});           
{\bf ESWW: }Engels et al.\,(\cite{eng85});       
{\bf FC99: }Forster et al.\,(\cite{for99});          
{\bf HTS:  }Hu et al.\,(\cite{hu94});                 
{\bf L89:  }Likkel\,(\cite{lik89});                   
{\bf L97:  }Lewis\,(\cite{lew97});                    
{\bf LC96: }te Lintel Hekkert et al.\,(\cite{tel96});
{\bf LDL:  }Lewis et al.\,(\cite{lew95});             
{\bf NB92: }Nedoluha et al.\,(\cite{ned92});         
{\bf NHWA: }Norris et al.\,(\cite{nor84});           
{\bf OWMS: }Olnon et al.\,(\cite{oln80});            
{\bf PPT:  }Payne et al.\,(\cite{pay88});             
{\bf SG04: }Szymczak et al.\,(\cite{szy04b});         
{\bf SHH:  }Slootmaker et al.\,(\cite{slo85});        
{\bf STZW: }Silverglate et al.\,(\cite{sil79});      
\end{list}
\end{table}

\section{Summary}
\label{summary}

The paper estimated (pseudo) OH 1612\,MHz maser pumprates or pumprate ranges for all OH/IR sources with at least one of the two OH 1612\,MHz maser IR pumping lines detected by ISO and all sources with ISO spectral observations at both pumping-line wavelengths (non-detections of both pumping lines). The correlation between (pseudo) maser pumprate or pumprate distribution and maser source types reveals three facts: (1) OH 1612\,MHz masers in late type stars show nearly fixed radiative pumprate, which agrees to the general thought that they are mainly radiatively pumped and hence the IR pumping-line flux should be proportional to the maser flux; (2) OH 1612\,MHz masers occurring in interstellar material show very small and largely varying (pseudo) radiative pumprates, which should be explained by four alternative scenarios or their combinations: absorption by quiet OH cloudlets, different OH maser pump mechanism, competitive gain between mainline and 1612\,MHz masers, anisotropy of the maser emission; (3) it is very likely that post-AGB stars, OH-PPNe and OH-PNe have their OH maser pumprates between that of  typical stellar and interstellar OH masers.

\appendix{Appendix: \\\\Line width of a doublet in low resolution spectrum}

Generally, the width of a line feature depends on many factors, such as natural width of energy levels, gas temperature, turbulence distribution, global expansion or contraction velocity distribution, instrumental resolution and all kinds of noise. But in this paper, the very low spectral resolution and the doublet nature of the line features allow one to only consider the instrumental resolution and separation of the doublet components when estimating the line width.

The 34.6 and 53.3\,$\mu$m OH absorption lines are both quadruple lines. But the four sub-lines of each feature can usually be divided into two groups, with two sub-lines in each group. The two sub-lines in each group is much closer to each other than to the lines in the other group. For 34.6\,$\mu$m line, the distance between the two sub-lines in each group is not larger than 0.0001\,$\mu$m, while the distance between the two groups is about 0.026\,$\mu$m. Further more, the widths of the four sub-lines are mainly controlled by the expansion velocity of the circumstellar shell from which it arises, i.e., typically equal to about 20\,km/s (equivalent to 0.0023\,$\mu$m at 34.6\,$\mu$m). Therefore, the 34.6\,$\mu$m quadruple can be considered as a doublet with two narrow components. For 53.3\,$\mu$m, the situation is similar. The four sub-lines can be divided into two groups with the separation of the two sub-lines within each group (not larger than 0.0004\,$\mu$m) much smaller than the distance between the two groups (about 0.09\,$\mu$m). Therefore it can also be considered as a doublet with narrow components.

The resolution of the ISO SWS/LWS is low. The SWS resolving power varies from 400 to 2250 (Although Fabry-P${\acute e}$rot (FP) spectra can attain a resolution of 30000, none of the spectra that need to estimate their line width in the paper were obtained by FP), which is equivalent to resolution varying from 0.112 to 0.041\,$\mu$m. The LWS resolution varies from 0.035 to 0.28\,$\mu$m. Comparing the resolution with the widths of each component of the 34.6 and 53.3\,$\mu$m doublets ($\approx  $0.0023\,$\mu$m, as stated above), it is seen that the components can be approximated as infinitely narrow lines when convolving with the ISO instrumental profiles (assumed as gaussian profile with the FWHM equals to spectral resolution). Thus, convolving the gaussian instrumental profile with the doublet, one can get a double-gaussian profile in which the two gaussian components peak at the wavelengths of the two doublet components and the FWHMs of both gaussian components equal to the spectral resolution. Assuming the strengths of the two doublet components are equal, the FWHM of the double-gaussian line profile can be approximately (not accurately!) expressed as:
$$FWHM = 0.026 \,\mu{\rm m} + R$$ 
for 34.6\,$\mu$m line; and
$$FWHM = 0.09 \,\mu{\rm m} + R$$ 
for 53.3\,$\mu$m line, where R is the spectral resolution expressed in unit $\mu$m and 0.026\,$\mu$m and 0.09\,$\mu$m are the separation between the two components of the 34.6 and 53.3$\mu$m doublets respectively.

\ack{I thank (an) anonymous referee(s) who suggested important improvements for the work in this paper. I am also grateful to Dr. Perter te Lintel Hekkert for kindly supplying additional information on the OH 1612\,MHz maser observations. This work is supported by  the Chinese National Science Foundation under Grant No. 10073018, the Chinese Academy of Sciences Foundation under Grant KJCX2-SW-T06 and the Yunnan Natural Science Fund (2002A0021Q).
}

\end{document}